# Investigating America Online Instant Messaging Application: Data Remnants on Windows 8.1 Client Machine

Teing Yee Yang [a], Ali Dehghantanha [b], Kim-Kwang Raymond Choo [c], Zaiton Muda [a]

[a] Department of Computer Science, Faculty of Computer Science and Information Technology, Universiti Putra Malaysia, UPM Serdang, Selangor, Malaysia

[b] The School of Computing, Science & Engineering, Newton Building, University of Salford, Salford, Greater Manchester, United Kingdom

[c] Information Assurance Research Group, Advanced Computing Research Centre, University of South Australia, Mawson Lakes Campus, Mawson Lakes Boulevard, Mawson Lakes, South Australia, Australia

**Abstract**

Instant messaging applications (apps) are one potential source of evidence in a criminal investigation or a civil litigation. To ensure the most effective collection of evidence, it is vital for forensic practitioners to possess an up-to-date knowledge about artefacts of forensic interest from various instant messaging apps. Hence, in this chapter, we study America Online Instant Messenger (version 7.14.5.8) with the aims of contributing to an in-depth understanding of the types of terrestrial artefacts that are likely to remain after the use of instant messaging services and app on Windows 8.1 devices. Potential artefacts identified during the research include data relating to the installation or uninstallation, log-in and log-off information, contact lists, conversations, and transferred files.

## 1. Introduction

Instant messaging (IM) is popular on both traditional computing devices (e.g., personal computers and laptops) and smart mobile devices (e.g. Android devices and iOS devices), as it allows users to exchange information with their peers in real-time using text messaging, voice and video messaging, and file sharing (e.g. pictures and videos). The number of worldwide IM accounts in 2015 is reportedly over 3.2 billion and is expected to increase to more than 3.8 billion by the end of 2019 (Radicati Team, 2015).

Similar to other popular consumer technologies, IM services have been exploited to commit frauds and scams (Timoney, 2014; Meyers, 2014; City of London Police, 2015; Farnden, 2015), dissemination of malware (Daryabar et al., 2011, 2013a; Mohtasebi et al., 2011b; Damshenas et al., 2013; Help Net Security, 2014), grooming children online with the purpose of sexual exploitation (Barnes, 2014; Godfrey, 2014; McCallum, 2013; FBI, 2015), and other criminal activities (Mohtasebi et al., 2011c; Dezfoli et al., 2013; Ganji et al., 2013; Daryabar et al., 2013b). The chat logs may provide information of evidential value to forensic practitioners (Norouzi et al., 2015; Ali, 2015), which can often be used to reconstruct events or reveal information such as a suspect's physical location, true identity, transactional information, incriminating conversations, and victim's information (e.g. email addresses and bank account numbers) (NCJRS, 2007).

Due to the increased user privacy requirements (Aminnezhad et al., 2012; Barghuthi and Said, 2013; Azfar et al., 2015a) and demands for data security, it is increasingly challenging to collect evidential data from IM service providers - similar to the trend of users storing data in a cloud platform (Shariati et al., 2015a; 2015b). The data are often protected using proprietary protocols and strong encryption, compounding the challenge to collect meaningful information from external networks (Damshenas et al., 2014). Even if the artefacts could be identified, collecting data from a multi-tenancy cloud environment may breach the data privacy policies of the IM service providers (ENISA, 2012; Dehghantanha and Franke, 2014). These challenges are compounded by cross-jurisdictional investigations, which may prohibit cross-border transfer of information (NCJRS, 2007; Damshenas et al., 2012).

Depending on the IM app in use, the end user device is often a potential source for the recovery of IM artefacts (Carvey, 2004; Dickson 2006a, 2006b, 2006c, 2007; Reust, 2006; Van Dongen, 2007a, 2007b; Parvez et al., 2011; Mohtasebi et al., 2011a; Ibrahim et al., 2012; Yusoff et al., 2014a, 2014b; Norouzi, 2015; Azfar et al., 2015b; Do et al., 2015; Immanuel, 2015; Leom, 2015; Damshenas et al., 2015; Talebi et al., 2015). Such terrestrial artefacts can be useful in establishing whether a suspect has a direct connection to a crime (Mohtasebi and Dehghantanha, 2013). While a practitioner should be cognisant of techniques of digital forensics, it is just as important to maintain an up-to-date understanding of the potential artefacts that could be forensically recovered from different types of IM products.

In this chapter, we seek to identify the potential terrestrial artefacts that may remain after the use of the popular America Online Instant Messenger desktop application (AIM) version 7.5.14.8 (AIM 7) on a Windows 8.1 client machine. Similar to the approaches of Quick and Choo (2013a, 2013b, 2014), we seek to answer the following questions in this research

1. What data remains on a Windows 8.1 device and their locations on a hard drive after a user has used AIM 7?
2. What data remains in Random Access Memory (RAM) after a user has used AIM 7 on a Windows 8.1 device?
3. What data can be seen in network traffic?

Findings from this research will contribute to the forensic community's understanding of the types of terrestrial artefacts that are likely to remain after the use of IM applications on devices running Windows 8.1 – the main contribution of this paper.

The structure of this chapter is as follows. Section 2 describes background and related work. Section 3 outlines the research methodology, including the framework and experiment environment and setup. In Sections 4 and 5, we present our findings from Windows 8.1, and conclude the paper as well as outlining future research areas.

## 2. Related Work

Dickson (2006a) investigated AIM 5.5 on a Windows XP machine and detected records of the contact lists and transferred files from registry, user settings, and other application-specific files on the hard disk. By searching for the suspect's screen name, Dickson (2006a) was able to reconstruct portion of the conversation texts from the unstructured datasets such as memory dumps, slack space, free space, and swap files in plain text and Hyper Text Markup Language (HTML), even without the chat logging. Reust (2006) concluded that the artefacts may not be of evidential value without timestamp information.

Kiley et al. (2008) approached evidence extraction of AIM Express (web application) and only recovered artefacts of the contact lists and conversations from the memory dump and hard disk's free space. However, the keyword search is only limited to unique phrases predefined in the experimental setup ('bannnnanas', 'weirdtheme', 'this is a space' etc.). A study of such may not be useful for evidence acquisition without prior knowledge of the incriminating conversations. Gladyshev and Almansoori (2010) proposed a method for acquiring known AIM conversation fragments from the memory dumps of Apple Mac computers. The researchers identified that the conversation fragments could be generally classified into 4 different formats, all of which held corresponding screen names, conversation texts, and timestamp information.

Yasin et al. (2012) examined the AIM protocol (and other IM protocols) for Digsby IM aggregator application and determined that the RC4 decryption key (for the Digsby password) is the SHA-1 hash of the concatenated system product identification (ID), install date, and Digsby client ID. In the subsequent studies, Yasin et al. (2013, 2014) examined the artefacts from memory dumps, swap files, and slack space and were able to recover complete conversation sessions from the unstructured datasets. Husain and Sridhar (2011) focused on logical acquisition of the AIM (version 2.0.2.4) artefacts for iOS and managed to retrieve forensically relevant information (i.e., records of conversations, user accounts, plain text password, and buddy list) from the iTunes backup of a non-jailbroken iPhone 3GS in Apple Property List (PLIST) format.

To the best of our knowledge, there is no published forensic research that studies the newer AIM client application on computer desktops running newer OS (i.e., Windows 8 or later) – a gap that this paper aims to contribute towards.

## 3. Research methodology

The examination procedure of this research was derived from the four-stage digital forensic framework of McKemmish (1999), which are identification of digital evidence, preservation of digital evidence, analysis, and presentation. The purpose is to enable acquisition of realistic data similar to that found in real world investigations. This paper mainly focuses on the analysis stage, although we also briefly discuss the evidence source identification, preservation, and presentation to demonstrate how this research can be applied in practice.

The first step of the experiment involved creation of four (4) fictional accounts to play the role of suspects and victims in this research – see Table 1. The IM accounts were assigned with a unique 'display icon' and username which was not used within the respective IM apps and Windows operating system. This eases identification of the user roles. Next was to create the test environments for the suspects and the victims, which consisted two (2) control base VMware Workstations (VMs) version 9.0.0 build 812388 running Windows 8.1 Professional (Service Pack 1, 64 bit, build 9600). As explained by Quick and Choo (2013a, 2013b, 2014), using physical hardware to undertake setup, erasing, copying, and re-installing would have been an onerous exercise. Moreover, a virtual machine allows room for error by enabling the test environment to be reverted to a restore point should the results are unfavourable. The workstations were configured with the minimal space (2GB of physical memory and 20GB hard drive space) in order to reduce the time required to analyse the considerable amounts of snapshots in the latter stage.

Table 1: User details for IM experiments.

| Username | Email | Role |
|---|---|---|
| Suspect | UC3F1211FC@gmail.com | Suspect |
| Victim | UC3F1211FC2@gmail.com | Victim 1 |
| VictimTwo | victimtwo@aol.com | Victim 2 |
| VictimThree | victimthree@aol.com | Victim 3 |

In the third step, we conducted a predefined set of activities to simulate various real world scenarios of using the apps on each workstation/test environment. The base assumptions are that the practitioner encounters a live system running Microsoft Windows 8.1 in a typical home environment. Similar to the approaches of Quick and Choo (2013a, 2013b, 2014), the 3111th email message of the University of California (UC) Berkeley Enron email dataset (downloaded from *http://bailando.sims.berkeley.edu/enron_email.html* on 24th September 2014) was used to create the sample files and saved as SuspectToVictim.rtf, SuspectToVictim.txt, SuspectToVictim.docx, SuspectToVictim.jpg, SuspectToVictim.zip, SuspectToVictim.jpg (printscreen), VictimToSuspect.rtf, VictimToSuspect.txt, VictimToSuspect.docx, VictimToSuspect.jpg (printscreen), VictimToSuspect.zip, and VictimToSuspect.jpg to simulate the transferring and receiving of files of different formats using the IM apps. As the filenames suggest, the 'SuspectToVictim' (and 'VictimToSuspect') files were placed on the suspect's workstation (and victims' workstations respectively) and subsequently transferred to the victims' workstations (and suspect’s workstation respectively). The Enron dataset was also used to create four (4) additional files, namely SUCCESS.dat, REJECTED.dat, ABORTED.dat, and BLOCKED.dat to examine the artefacts in relation to successful (accepted), rejected, aborted, and blocked file transfers.

The experiments were predominantly undertaken in NATed (where NAT stands for Network Address Translation) network environment and without firewall outbound restriction to represent a typical IM situation. Wireshark was deployed on the host machine to capture the network traffic from the suspect’s workstation for each scenario. After each experiment was carried out, we saved a copy of the network capture file in .PCAP format, and acquired a bit-stream (dd) image of the virtual memory (.VMEM) file prior to shutdown. We then took a snapshot of each workstation after being shutdown and made a forensic copy of the virtual disk (.VMDK) file in Encase Evidence (E01) format. This resulted in the creation of ten (10) snapshots (each for each environment) as highlighted in Table 2, Fig. 1, and Fig. 2. The decision to instantiate the physical memory dumps and hard disks with the virtual disk and memory files was to prevent the datasets from being adulterated with the use of memory/image acquisition tools (Quick and Choo, 2013a, 2013b, 2014).

Table 2: Details of VM snapshots created for AIM 7 investigations on Window 8.1.

| Snapshot | Description |
|---|---|
| **1.0 Base-Snapshot** | A control base snapshot was made to create the control media to determine changes from each IM scenario. |
| **A1.1 Install-Snapshot** | Using a duplicate copy of the control base snapshot (1.0), we downloaded and subsequently installed the AIM client software version 7.5.14.8 from https://www.aim.com/download#windows. |
| **A1.1.1 Login-Snapshot** | A snapshot was created of the install snapshot (A1.1) to analyse the process of logging in AIM 7 on a Windows 8.1 machine. The options “Remember Me”, “Remember My Password”, and “Automatically Sign Me In” were checked as is the case by default. |
| **A1.1.2 Buddy List-Snapshot** | A second snapshot was made of the install snapshot (A1.1) to examine the process of adding contacts and saving the Buddy List on a Windows 8.1 machine. Using the suspect’s account, we added the victim accounts ‘VictimTwo’, ‘uf3f1211fc2@gmail.com’, and ‘VictimThree’ to three separate groups, namely ‘Buddies’, family’, and ‘Group 1’ (custom group) respectively. We then assigned a friendly name (i.e., ‘Phantom Friend 1’, ‘Victim’, and ‘Phantom Buddy 1’ respectively) for each of the victim accounts. Finally, we saved the Buddy List as ‘savedbuddylist.blt’ on the desktop of the suspect’s workstation. |
| **A1.1.3 Conversations/File transfers-Snapshot** | A third copy of the install snapshot (A1.1) was made to undertake mock conversations and file transfers between two chat participants using the default settings (without IM logging). |
| **A1.1.4 IM Logs-Snapshot** | A fourth copy of the install snapshot (A1.1) was made to examine the conversation and file transfer artefacts of the IM logging, which was explicitly enabled prior to carrying out the experiment. |
| **A1.1.4.1 Uninstall-Snapshot** | A snapshot was created of the IM Logs snapshot (A1.1.4) to examine the artefacts left behind after uninstalling AIM 7 on a Windows 8.1 client machine. Uninstallation was undertaken using the ‘Uninstall AIM’ executable file. |
| **A1.1.5 Chat room-Snapshot** | A fifth snapshot was created of the install snapshot (A1.1) to examine the group conversation artefacts of AIM 7. The suspect’s account was used to create a chat room namely ‘devi’ and subsequently add all the victims into the chat room for group |
| | conversations. |
| **A1.1.6 Buddy Icon-Snapshot** | Another snapshot was made of the install snapshot (A1.1) to examine the user image ('buddy icon' in the context of AIM) artefacts of AIM 7. |
| **A1.1.7 What is Happening-Snapshot** | A final snapshot was made of the install snapshot (A1.1) to examine the process of creating personal message ('what is happening' message in the context of AIM) in AIM 7. |

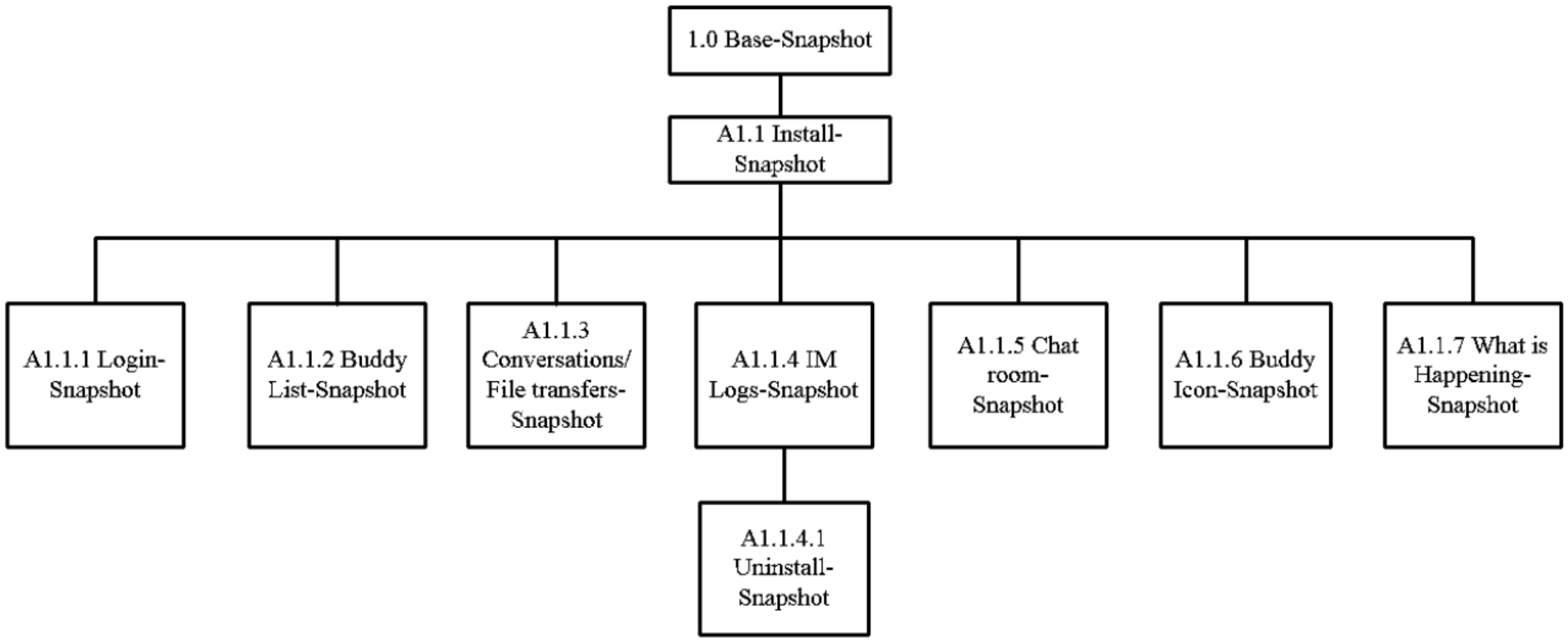

Figure 1: VM snapshots created for AIM experiments.

The final step of this research was to analyse the datasets using a range of forensically recognised tools (as highlighted in Table 3) and present the findings. Both indexed and non-indexed as well as Unicode and non-Unicode string searches were included as part of the evidence searches. The experiments were repeated at least thrice (at different dates) to ensure consistency of findings.

Table 3: Tools used for IM analysis on Windows 8.1.

| Tool | Usage |
|---|---|
| **FTK Imager Version 3.2.0.0** | To create forensic images for the .VMDK files. |
| **Autopsy 3.1.1** | To parse the file system, produce directory listings, as well as extracting or analysing stored files, browsing history, 'NTUSER.dat' registry files (using the RegRipper plugin), 'pagefile.sys' Windows swap file, and unallocated spaces located within the forensics images of VMDK files. |
| **HxD Version 1.7.7.0** | To conduct keyword searches in the unstructured datasets. |
| **Volatility 2.4** | To analyse the running processes (using the 'pslist' function), network statistics (using the 'netscan' function), and detecting the location of a string (using the 'yarascan' function) recorded in the physical memory dumps. |
| **Photorec 7.0** | To data carve the unstructured datasets. |
| **SQLite Browser Version 3.4.0** | To view the contents of SQLite database files. |
| **Wireshark version 1.10.1** | To analyse the network traffics. |
| **Network Miner version 1.6.1** | To analyse and data carve the network files. |
| **Whois command** | To determine the registration information of the IP addresses. |
| **Nirsoft Web Browser Passview 1.19.1** | To recover the credential details stored within web browsers. |
| **Nirsoft cache viewer, ChromeCacheView 1.56, MozillaCacheView 1.62, IECacheView 1.53** | To analyse the web browsing cache. |
| **BrowsingHistoryView v.1.60** | To analyse the web browsing history. |
| **Thumbcacheviewer Version 1.0.2.7** | To examine the Windows thumbnail cache. |
| **Windows Event Viewer Version 1.0** | To view the Windows event logs. |
| **Windows File Analyser 2.6.0.0** | To analyse the Windows prefetch and link files. |
| **NTFS Log Tracker V1.2** | To parse and analyse the $LogFile, $MFT, and $UsnJrnl New Technology File System (NTFS) files. |

## 4. AIM forensics

AIM is a free messaging application first launched in May 1997 and the Windows version was released on 31st May 2012 (Old Apps, n.d.). The users are required to sign up for an account (nominally a 'screen name' in the context of AIM) in order to use the service. AIM reportedly uses a proprietary communication protocol called the Open System for Communication in Realtime (OSCAR), which has been the subject of various reverse-engineering attempts (CMU, 2006).

AIM keeps track of the contacts in the Buddy List (the term 'Buddy' is an AIM term for contact), which can be saved as a text file and exported in .BLT format. Although there are default groups for buddies, family, and co-workers, custom groups can be added. There is also a special 'Recent Buddies' group that holds the screen names with whom the user recently contacted (AOL, 2015). The Buddy List also holds user-specific preferences for the contacts (or screen names added), such as nicknames (friendly names in the context of AIM), display of IMs, notification sounds, and other relevant settings (AOL, 2015).

In this section, we present results of our investigation of artefacts left behind after the use of AIM version 7.5.14.8 on a Windows 8.1 client machine, such as usernames, passwords, profile pictures, contact lists, away messages, personal messages, text of conversations, transferred files, and the associated the timestamps.

### 4.1. Installation of the AIM 7 Client Application

Analysis of the directory listing determined that the installation and application folders can be located at *%Program Files%\AIM* and *%AppData%\Local\AIM*, respectively. We also located link files (shortcuts) for the loader file (*%Program Files (x86)%\AIM\aim.exe*) at *%Desktop%\AIM.lnk* and *%AppData%\Roaming\Microsoft\Internet Explorer\Quick Launch\AIM.lnk*, which held timestamp records that reflected the install (creation) and last accessed times.

Examinations of the prefetch files (stored in *%SystemRoot%\Prefetch*) revealed five (5) prefetch files associated with the AIM client application, namely ‘AIM.EXE.pf’, ‘AIMINST.EXE.pf’, ‘AIMLAN~1.EXE.pf’, ‘SETUP.EXE.pf’, and ‘INSTALL_AIM.EXE.pf’. The prefetch files provided additional information such as the number of times the application has been loaded, last run time, and other associated timestamps. Analysis of the thumbnail cache in *%AppData%\Local\Microsoft\Windows\Explorer* located thumbnail images for the loader file, installer, and client application’s logo, which may be indicative of recent AIM usage.

Looking through the registry entries, it was observed that the installation can be ascertained from the presence of the *HKEY_LOCAL_MACHINE\SOFTWARE\Wow6432Node\America Online* and *HKEY_LOCAL_MACHINE\SOFTWARE\Wow6432Node\AOL* registry hives, but it appears that only the installation metadata i.e., directory paths for the loader, diagnostic, and shortcut files can be recovered from the registry. Similar to any other Windows applications, when the AIM client was configured to run automatically whenever the Windows starting, we were able to locate the last run time for the client application in *Software\Microsoft\Windows\CurrentVersion\Run*. Figure 2 illustrates that the entry could be differentiated by its verb[1].

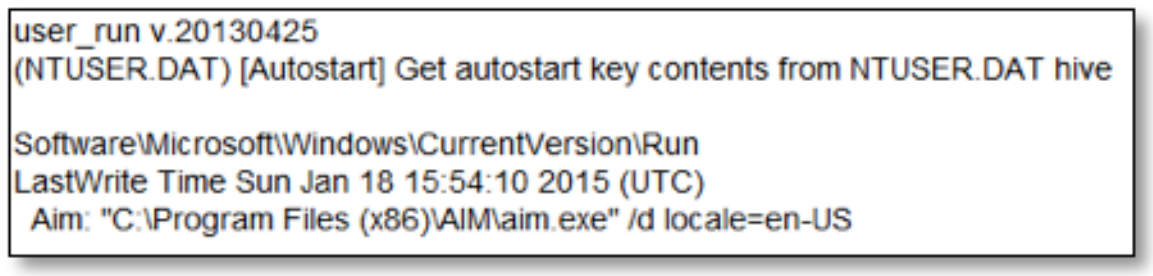
```
user_run v.20130425
(NTUSER.DAT) [Autostart] Get autostart key contents from NTUSER.DAT hive

Software\Microsoft\Windows\CurrentVersion\Run
LastWrite Time Sun Jan 18 15:54:10 2015 (UTC)
  Aim: "C:\Program Files (x86)\AIM\aim.exe" /d locale=en-US
```

Figure 2: Last run time recorded in the ‘Run’ registry key.

Analysis of the *Software\Microsoft\Windows\CurrentVersion\Explorer\ComDlg32* registry revealed the last accessed time for the loader file in the ‘CIDSizeMRU’, ‘OpenSavePidlMRU,’ and ‘LastVisitedPidlMRU’ registry subkeys (see Figure 3), where MRU is the abbreviation for Most-Recently-Used. The findings suggested that the AIM client was recently used, had been opened or saved within a Windows shell dialog box, and was used to open the files documented in the ‘OpenSaveMRU’ subkey, respectively (Carvey, 2004). Additionally, we also located references to the directory paths and last access times for the loader and link files in *Software\Microsoft\CurrentVersion\Explorer\UserAssist*, indicating that the AIM client was opened frequently on the machine under investigation (Nirsoft, 2015).

[1] A verb is a text string used by the Windows Shell to define the command on the shortcut menu of a file (Microsoft Inc., 2015b).

```
comdlg32 v.20121008

Software\Microsoft\Windows\CurrentVersion\Explorer\ComDlg32
LastWrite Time Sun Jan 18 16:03:39 2015 (UTC)
CIDSizeMRU
LastWrite: Sun Jan 18 16:03:39 2015
Note: All value names are listed in MRUListEx order.

  aim.exe

LastVisitedPidlMRU
LastWrite: Sun Jan 18 16:03:39 2015
Note: All value names are listed in MRUListEx order.

  aim.exe - My Computer\Unknown Type (0x2e)
```

Figure 3: Last accessed time recorded in the 'ComDlg32' registry key.

Examination of the running processes using the 'pslist' function of Volatility revealed the process identifier (PID), parent process identifiers (PPID), as well as process initiation and termination times from the memory dump; an excerpt is shown in Figure 4. The PID can prove useful for correlating data associated with the client application during further analysis (i.e., using the 'Yarascan' function of Volatility).

| Offset(V) | Name | PID | PPID | Thds | Hnds | Sess | Wow64 | Start | Exit |
|---|---|---|---|---|---|---|---|---|---|
| 0xffffe00000486900 | explorer.exe | 2988 | 2960 | 51 | 0 | 1 | 0 | 2015-01-19 03:48:22 UTC+0000 | |
| 0xffffe00000657900 | aim.exe | 2704 | 2988 | 0 | | 1 | 0 | 2015-01-19 03:48:46 UTC+0000 | 2015-01-19 04:06:37 UTC+0000 |

Figure 4: An excerpt of the 'pslist' output for the AIM client.

## 4.2. Logins

By default, in AIM 7 'Remember Me', 'Remember My Password', and 'Automatically Sign Me In' options are enabled. The 'Remember My Password' option stores the login credentials using Blowfish encryption (along with base 64 encoding) in the aimx.bin file located in the application folder. From a digital forensics perspective, this option can be beneficial to provide investigator's access to the suspect's online AIM account from the computer or forensic image under investigation without needing the user password. Moreover, the option can also provide a practitioner the opportunity to recover the credential details using AIM Password Decryptor (Talekar, 2011). Alternatively, it was observed that the user's screen name/login email can be located in cache files such as *%AppData%\Local\Microsoft\Windows\INetCache\IE\<Cache ID>\AIM_UAC_v2.htm* and *%AppData%\Roaming\acccore\caches\users\<User's Screen Name>\buddyicon\bartIDs_devformat_01*.

An inspection of the network log (*%AppData%\Local\AIM\Logs\network_log_#.txt*) located references to the host's external IP addresses, prefixed by 'host address' i.e., "00:26.29 Connection 039456E8: host address 152.163.9.73". Pairing the corresponding IP addresses/screen name with the relevant timestamps will allow a practitioner to correlate any external data that might have been obtained from an ISP or other external provider. Analysis of the memory dump was able to recover copies of the login password in Unicode string format, but no obvious pattern could be located to enable future searches.

When the logins occurred, it was observed that the host first established a session with Cyberlink Internet Services AG (i.e., IP address 62.12.173.139 on port 443) for key authentication and then to the America Online (AOL) server (i.e., IP address 64.12.104.89), but we were unable to identify the purpose of the latter due to lack of information from the URL as well as encrypted traffic. The next IP addresses accessed were 205.188.87.7 and 205.188.98.4 (in our research) on port 80 to retrieve the certification revocation lists from the AOL servers. Afterwards, the host accessed the login server (i.e., IP address 207.200.74.12 on port 443 with URL referencing 'my.screenname.aol.com') for login authentication. A search for the user's screen name returned multiple matches in the HTTP 'referer' request header for the Atwola advertisement cookie tracking server (i.e., IP addresses 64.12.96.217 and 207.200.74.71) in the form of 'http://www.aim.com/redirects/inclient/AIM_UAC_v2.adp?locale=en-US&magic=93321503&width=180&height=150&sn=<Screen name>', indicating that the user's screen name can be potentially recovered from the network captures using the link identified. Table 4 lists the network information observed in our research.

Table 4: Network information observed for the AIM client.

| Registered owner | IP address(es) | URL(s) observed |
|---|---|---|
| **Cyberlink Internet Services AG** | 62.12.173.139 | Kdc-aim.egslb.aol.com, Kdc.uas.aol.com |
| **AOL. Inc.** | 64.12.104.89 | bos-m016a-new-rdr2.blue.aol.com |
| **AOL. Inc.** | 149.174.110.118 | www.aol.com |
| **AOL. Inc.** | 205.188.14.120 | ars.oscar.aol.com (Windows) |
| **AOL. Inc.** | 205.188.87.7 | crl.egslb.aol.com, crl.aol.com |
| **AOL. Inc.** | 205.188.88.125 | abapi.abweb.aol.com |
| **AOL. Inc.** | 205.188.98.4 | ocsp.egslb.aol.com, ocsp.web.aol.com |
| **AOL. Inc.** | 207.200.74.66 | www.aim.com |
| | 199.7.52.72 | ocsp.verisign.net, ocsp.verisign.com |
| **AOL. Inc.** | 207.200.74.12 | my.screenname.aol.com.aol.akadns.net, my.screenname.aol.com |
| **AOL. Inc.** | 64.12.96.217, 207.200.74.71 | at.atwola.com |

Reconstructing the network captures using Netminer, we only recovered certificates that were used to authenticate the HTTPS sites, HTML documents from the HTTP sites, as well as JavaScript and XML files for the advertisement cookies. Although most of the network traffics were encrypted and the credential details were not recovered, the IP addresses and URLs highlighted as part of our research may assist a practitioner in scoping and timelining the AIM activities undertaken by a suspect in future investigations. Alternatively, it was determined that the network information can be partly recovered from the memory dump using the 'netscan' function of Volatility, which includes the associated PID, process creation time (if any), and socket states.

### 4.3. Buddy Lists

Examinations of the directory listings only located the manually saved Buddy List, which could be easily differentiated by the extension .BLT. An inspection of the Buddy List saved in our experiment observed that the user's screen name was prefixed by 'screenName' in the 'User' tag. The 'Buddy' tag held details associated with the groups, each group name formed an opening and closing subtag (curly bracket) in the 'list' subtag. Within the group name subtags there were additional subtags that held the nicknames (friendly names in the context of AIM) and personal notes for the screen names added to the groups. However, it is noteworthy that the saving of group names as well as nicknames and personal notes for the contacts' screen names is subject to user-settings for the Buddy List.

A search for the suspect and the victim's screen names and group names such as 'Co-Workers', 'Family', and 'Recent Buddies' recovered copies of the saved Buddy List from the memory dump in plain text. It was also identified that the saved Buddy List can be differentiated from the header "43 6F 6E 66 69 67 20 7B" in hexadecimal or "Config.{" in American Standard Code for Information Interchange (ASCII) format, but there was no common structure to determine the footer, and hence data carving is not possible for the saved Buddy List. In the circumstance when the Buddy List was not manually saved, we could recover clear text copies of the screen names displayed in the Buddy List window (of the client application) from the memory dump and swap file. Figure 5 illustrates that a practitioner can potentially locate the screen names with whom a suspect had recently communicated following the group name 'Recent Buddies' in the unstructured datasets.

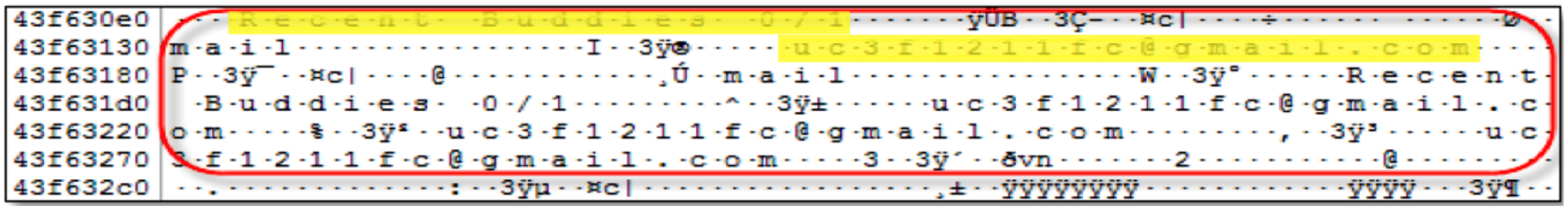

Figure 5: Remnants of recent buddy list recovered from RAM in plain text.

### 4.4. Conversations and Transferred Files

The downloaded files were saved in *%Desktop%* by default. Each of which was given an Alternate Data Stream (ADS) ZoneTransfer marker (ZoneID) with reading 'ZoneID=3', indicating that the files were downloaded from Internet zone (Microsoft Inc., 2015a). This suggests that a practitioner can potentially identify files associated with the download from files with ADS ZoneID=3 on the desktop. Inspection of the timestamps of the downloaded files revealed that the creation and accessed times were the times when the files were successfully downloaded on the recipient's machine, and the modified time of the receiving files on the recipient's machine remained unchanged from the original modified time of the files on the sender's machine. The last written time was retained from the original last written time recorded on the sender's machine.

When the transferred and downloaded files were opened, we located shortcuts for the transferred and downloaded in the recent documents directory (located at *%AppData%\Roaming\Microsoft\Windows\Recent*). The shortcuts can provide a potential alternative method for timelining the last modified, accessed, and creation times associated with the transferred or downloaded files. The findings also indicate records of the last accessed times can for the viewed files in the *Software\Microsoft*\Windows*CurrentVersion\Explorer\RecentDocs* (henceforth 'RecentDocs') registry key. Analysis of the thumbcache files (located in *%AppData%\Local\Microsoft\Windows\Explorer*) recovered thumbnail images for the transferred and downloaded PDF and image files, suggesting that thumbnail cache is a source for possible data associated with AIM use.

When the IM logging was not enabled, we only managed to recover artefacts of the conversations from unstructured datasets such as memory dump and swap file. The artefacts contained incoming and outgoing fragments of conversations texts, screen names of the user and the correspondent(s), away messages, filenames for the transferred/downloaded files, as well as the associated timestamp information in plain text and HTML; Figure 6 shows an example of the plain text remnants in Unicode string format. A search for the texts (from the remnants) using the 'Yarascan' function of Volatility determined that most of the artefacts were attributed to 'aim.exe' in the memory dump, indicating the texts were remnants from the AIM client application. The suspect's screen name can be a useful keyword for future searches. Our findings also revealed that filenames for the successful transferred files can be found prefixed by the terms 'Cool FileXfer'[2] as well as 'transfer complete' in the unstructured datasets; we hypothesise that the former was remnants from the payloads of the file transfer threads. Once the filenames are identified for the transferred/downladed files, the practitioner can correlate filenames with the directory listing, $LogFile, $MFT, $UsnJrnl, or other sources of relevance to identify file paths.

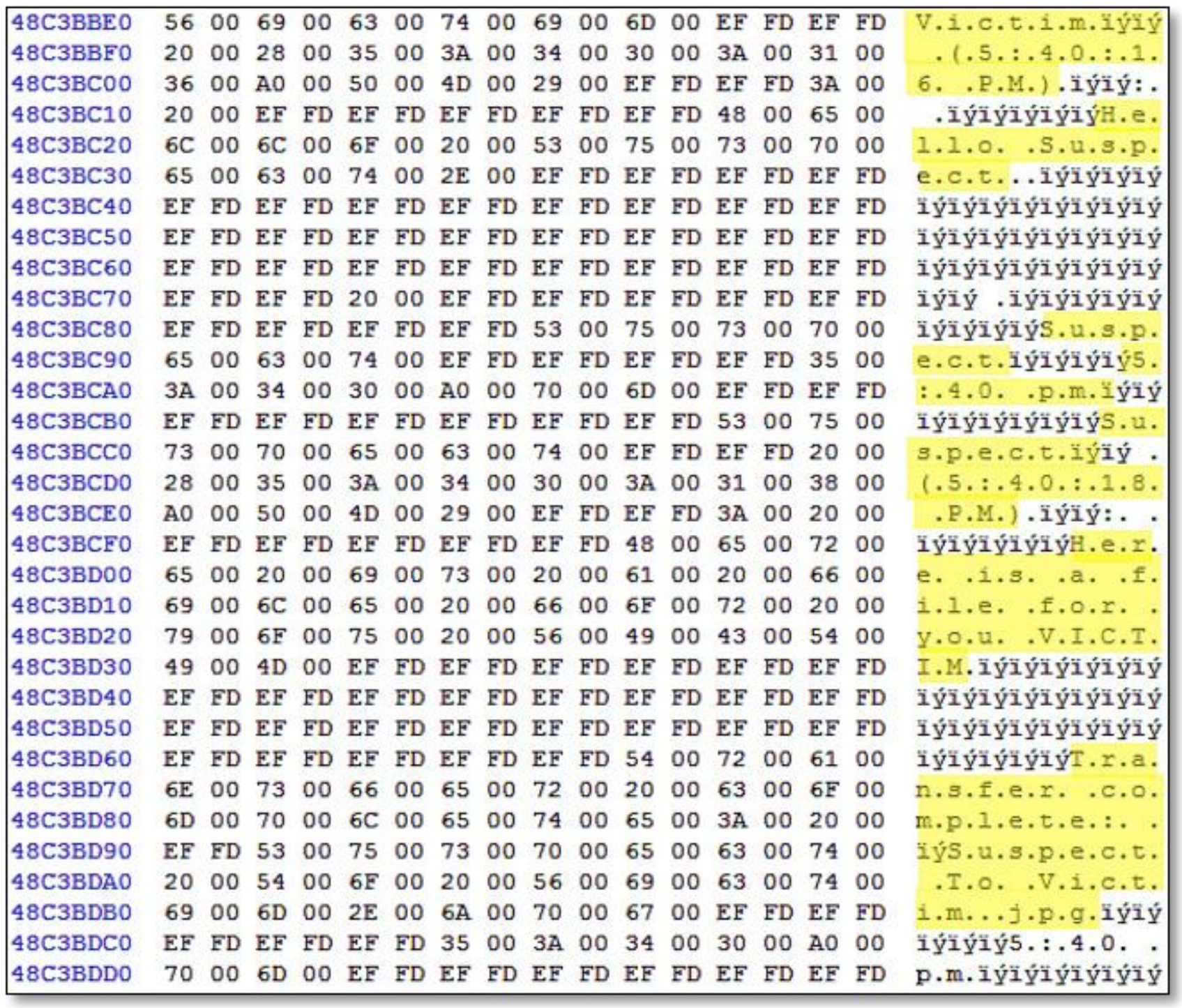

Figure 6: Remnants of AIM conversations recovered from suspect's RAM in plain text.

Undertaking data carving of the memory dump recovered copies of the transferred and downloaded files intact, with the exception of the .DAT files as there were no common header and footer information to determine the file structure. Alternatively, it was observed that contents of the transferred and downloaded files can be manually recovered from the memory dump in plain text. The artefacts can be useful in the circumstance when data carving is not applicable i.e., due to missing header and footer information. In a real world circumstance, the contents can be identified using a number of methods including directly from the user, hints from the incriminating IM conversations, or via the filename references recorded in the registry, directory listings, memory dump, and other unstructured datasets.

[2] 'Cool FileXfer' is the identification string for the OSCAR File Transfer 3 (OFT3) Header (CMU, 2006).

Analysis of the network traffic observed that the conversations were established with 64.12.104.* (see Table 4) on port 443. When the file transfers occurred, it was observed that the host established a direct TCP connection with the correspondents, and hence the IP addresses could be detected. In the case when direct communication was prohibited (CMU, 2006), the sessions were engaged with the proxy server i.e., IP address 205.188.14.120 in our research. OFT3 was seen as the carrying protocol for the both types (direct and proxied) of transfers, from which we managed to recover filenames for the transferred files, prefixed by 89-byte NULL Dummy block (CMU, 2006) in the respective headers. It was also observed that transfers that were completed could be differentiated from the header 'type' ($7^{th}$ to $8^{th}$ byte) given the value of '0x0204', indicating that the recipients had received the files successfully (CMU, 2006); an example is shown in Figure 7. The identification string 'Cool FileXfer' can be a suitable keyword for future searches.

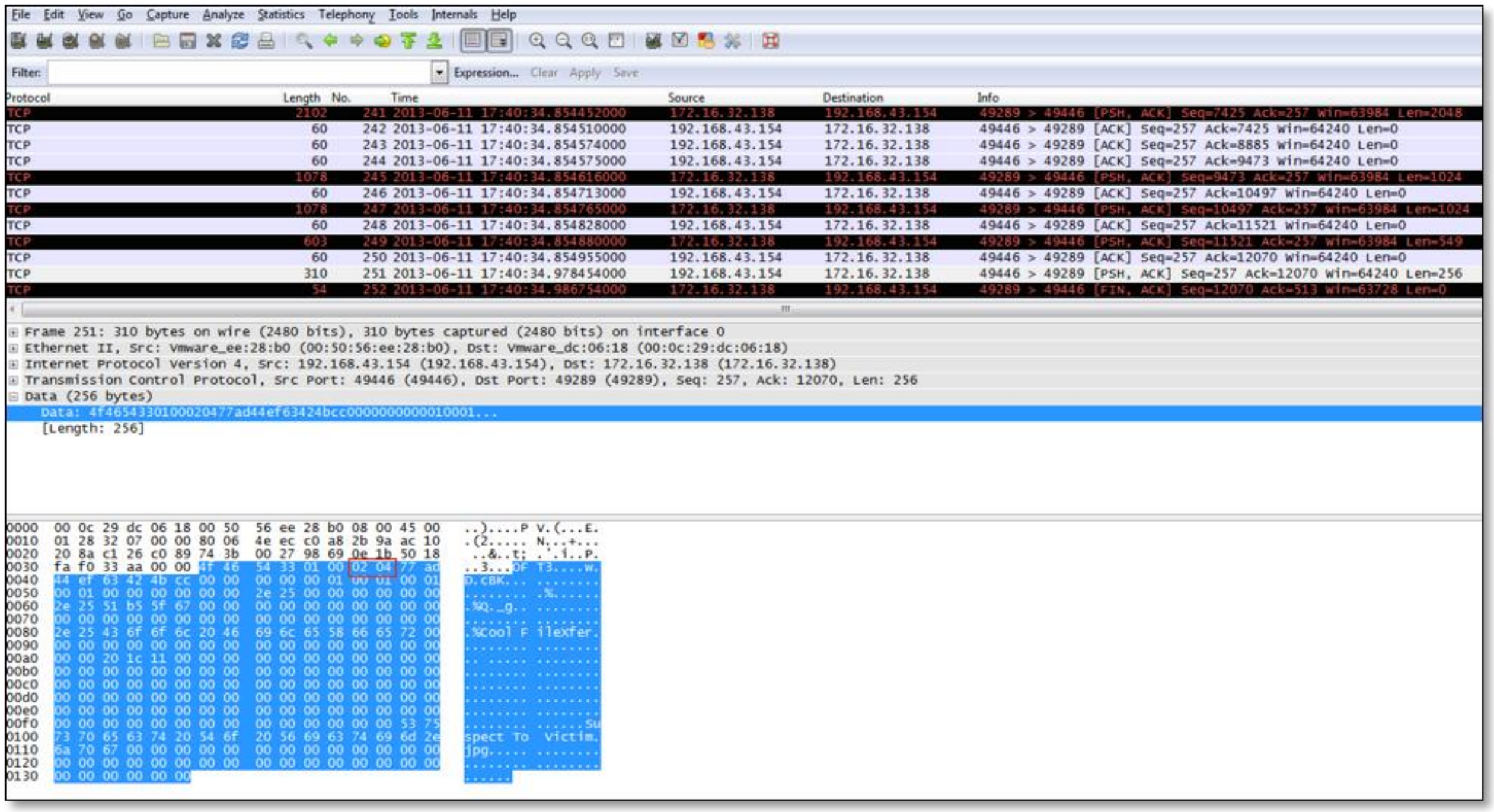

Figure 7: OFT3 file transfer Header for successful file transfer.

## 4.5. IM Logs

The IM logs were saved at *%Documents%\AIMLogger\[User's screen name]\IM Logs\[Correspondent's Screen Name].html* by default. As the filename suggests, the AIM client application created a new log for each screen name with whom the user had communicated. Analysis of the IM logs determined that incoming and outgoing text of conversations, away status, and the corresponding timestamp information can be recovered from the IM logs, but not for the away messages, file transfer statuses, and filenames for the transferred/downloaded files. Alternatively, it was determined that portion of the IM logs can be located in *%ProgramData%\Microsoft\Search\Data\Applications\Windows\Windows.edb* and *%ProgramData%\Microsoft\Search\Data\Applications\Windows\edb#####.log* (when a log file is nearing 1MB, a new log file, numbered sequentially, will be created) using the phrase ‘IM history with buddy’. The Windows.edb file is a database of the Windows Search service which provides search results, property caching, and content indexing for files, emails and other relevant data stored in the file system (Metz, n.d.).

Examinations of the unstructured datasets (i.e., memory dump, swap file, and unallocated space) identified that the IM Logs can be potentially carved using the header and footer value of “3C 3F 78 6D 6C 20 76 65 72 73 69 6F 6E 3D 22… 3C 2F 62 6F 64 79 3E 0D 0A 3C 2F 68 74 6D 6C 3E”, but the findings may be subject to software updates. A search for the suspect’s screen name and phrase ‘IM history with buddy’ produced matches to the IM logs in the unstructured datasets in plain text and HTML, indicating that copies of the IM logs can be extracted from unstructured datasets via keyword search. An inspection of the HTML coding indicated the corresponding screen names, conversation texts, timestamps, and other conversation information as detailed in Figure 8; the values ‘LOCAL’ and ‘REMOTE’ in the ‘class’ attribute reflect incoming and outgoing conversations respectively.

```
<tr><td class='time'>[DAY,MONTH,NUMBER OF DAY,YEAR]</td></tr>..<tr><td class='[LOCAL OR REMOTE]'>[IM SCREEN NAME];[(TIME IN HOUR:MINUTE:SECOND;AM OR PM])</td><td class='msg' width='[WIDTH]'><FONT face='[FONT TYPE]' size='[SIZE]' color='[COLOR CODE]'>[MESSAGE CONTENT]</FONT></td></tr>
```

Figure 8: Excerpt of HTML coding for the IM log.

### 4.6. Buddy Icons

A ‘buddy icon’ is an AIM term for user image. Artefacts of the user images (for the user and the correspondents) were only located in the thumbcache folder as well as unstructured datasets (i.e., memory dump, swap file, and unallocated space) in JPEG format, but there were no common structures to enable future identification of the user images. However, it was determined that a practitioner can retrieve the currently used image for a screen name from the Application Programming Interface (API) link *http://api.oscar.aol.com/expressions/get?f=native&type=buddyIcon&t=[Screen Name]*.

### 4.7. ‘What is Happening’ Messages

A ‘What is Happening’ message is an AIM term for personal message. This feature can be exploited in social engineering (i.e., scam) where a criminal broadcasts hoaxes or links to malware or phishing to lure potential victims. In our research, it was observed that artefacts of the personal messages could only be located in the memory dump, with no associated string that would enable future searches. However, it was identified that a practitioner may potentially retrieve the recently updated personal messages in relation to a screen name alongside the timestamp information from the Lifestream[3] profile via the link *http://lifestream.aol.com/[Screen Name]*.

### 4.8. Uninstallation of AIM Client Software

During uninstallation, an AIM user may choose to remove the application data and IM logs from the client device completely; both options were selected in our research. An analysis of the directory listing determined that only *%AppData%\Local\AIM* and *%AppData%\Local\AOL\AOLDiag* remained after uninstallation, but the folders were empty. The uninstallation process also created two uninstaller files (‘A~NSISu_’ and ‘B~NSISu_’) in *%AppData%\Local\Temp* as well a prefetch file (UNINST.EXE.pf), and the last accessed/run timestamp of which could indicate the uninstall time.

---

[3] Lifestream is an integrated feature in AIM Express, AIM 7 and AIM for Mac that keeps track of updates made by the user and the correspondents on AIM, Facebook, Twitter, and other social networking or IM platforms (AOL, 2014).

Inspecting the registry files, it was observed that only *HKCU\Software\America Online*, *HKLM\SOFTWARE\Wow6432Node\America Online*, and *HKLM\SOFTWARE\Wow6432Node\AOL* were retained, but the entries were removed. Other remnants (from memory dump, swap files, unallocated space, thumbcache, ‘RecentDocs’, ‘UserAssist’, ‘Run’, and ‘ComDIg32’ registry keys etc.) were unaffected by the uninstallation, but results may not be definitive.

## 5. Conclusion and Future Work

In this research, we discussed forensic analysis of artefacts left by AIM 7 on Windows 8.1 desktop client. In general, our results concluded the opposite from the previous studies (Dickson, 2006; Reust, 2006) based on registry, application data, contact and log files. One of the major changes is that the registry no longer holds caches of recent conversations and login credentials, suggesting that registry is no longer a source of potential data for AIM 7. Although there were no common structures to enable identification of the user images and personal messages on the desktop client, our findings indicate that a practitioner may potentially retrieve the latest user image and recently updated personal messages from the server using the corresponding links identified in our research. As with any other Windows client applications and OS versions, our examination of the system files such as shortcuts, event logs, thumbcache, $LogFile, $MFT, $UsnJrnl, as well as registry keys i.e., ‘RecentDocs’, ‘UserAssist’, ‘Run’, and ‘ComDig32’ revealed that additional timestamp and file path information can be recovered to support evidence found on the target system.

In the absence of the user manual saved buddy list and IM logs, it appears that artefacts of the contact lists (buddy lists) and conversations can only be recovered from memory dump in an unstructured form, although portion of the HTML coding can be recovered in some cases. This seems to agree with the findings of Dickson (2006) and Reust (2006). The fact that there was no clear text password in the memory dump should perhaps not be unsurprising since the credential information is securely encrypted in the aimx.bin file (Talekar, 2011). With the use of known file structures, we determined that it is possible to recover copies of the IM logs and transferred files from the memory dump intact, assuming that the data is not overwritten. Nevertheless, a practitioner must keep in mind that memory changes frequently according to users’ activities and will be wiped as soon as the system is shut down. Our experiments also suggested that a practitioner can potentially recover portion of the conversations and transferred files from the swap files and unallocated space, which are potentially resulted from remnants of the inactive memory pages swapped to the hard disk. Hence, a practitioner should not underestimate the importance of analysing physical memory dump, swap file, and unallocated space when undertaking forensic investigation in relation to the newer AIM client application.

Although most of the network traffics were encrypted and the credential information was not recovered, we contend that the IP addresses and URLs highlighted as part of our research may assist a practitioner in scoping the actions undertaken by a suspect on AIM, ensuring that relevant follow-up actions can be undertaken in a timely manner. Table 5 summarises the key artefacts located as part of our research. As the remnants were recovered with minimal space configuration in our research, we believe there will be a greater chance of remnants on a typically larger system.

Table 5: Summary of findings.

| **Source of evidence** | **Details** |
|---|---|
| **Registry branches of forensic interest** | • *HKEY_LOCAL_MACHINE\SOFTWARE\Wow6432Node\America Online*<br>• *HKEY_LOCAL_MACHINE\SOFTWARE\Wow6432Node\AOL*<br>• Software\Microsoft\CurrentVersion\Explorer\UserAssist<br>• Software\Microsoft\Windows\CurrentVersion\Run<br>• Software\Microsoft\Windows\CurrentVersion\Explorer\ComDlg32<br>• Software\Microsoft\Windows\CurrentVersion\Explorer\RecentDocs |
| **Files/folders of forensic interest** | • *%AppData%\Local\aimx.bin*<br>• *%AppData%\Local\AIM\Settings\[User's Screen Name]\settings.xml*<br>• *%AppData%\Local\Microsoft\Windows\INetCache\IE\<Cache id>\AIM_UAC_v2.htm*<br>• *%AppData%\Roaming\acccore\caches\users\<User's Screen Name>\buddyicon\bartIDs_devformat_01*<br>• .BLT buddy list file<br>• *%Documents%\AIMLogger\[User's screen name]\IM Logs\[Correspondent's Screen Name].html*<br>• *%ProgramData%\Microsoft\Search\Data\Applications\Windows\Windows*<br>• *%ProgramData%\Microsoft\Search\Data\Applications\Windows\edb#####.log* |
| **URLs of forensic interest** | • *http://api.oscar.aol.com/expressions/get?f=native&type=buddyIcon&t=[Screen Name]*<br>• *http://lifestream.aol.com/[Screen Name]* |
| **Prefetch files** | • AIM.EXE.pf<br>• AIMINST.EXE.pf<br>• AIMLAN~1.EXE.pf<br>• SETUP.EXE.pf<br>• INSTALL_AIM.EXE.pf |
| **Link files** | • The link file could be discerned from 'AIM.lnk'<br>• Located link files for the transferred or downloaded files in *%\AppData\Roaming \Microsoft\Windows\Recent* |
| **Thumbcache files** | • Icon images used by the client application<br>• Thumbnail images for the transferred or downloaded PDF and image files<br>• User images of the user and the contacts |
| **Memory dumps, swap files, and unallocated space** | • Copies of the files of forensic interest as well as transferred or downloaded files unencrypted<br>• Filename and path references for the files of forensic interest and transferred or downloaded files<br>• Unstructured screen names and conversation texts, which included the timestamp information<br>• Copies of the Buddy List appeared in the Buddy List windows<br>• The process name could be discerned from 'aim.exe' in the memory dump |
| **Network traffics** | • OFT3 header containing filenames for downloaded/transferred files and the transfer statuses<br>• Host and servers' IP addresses<br>• Host and correspondents' IP addresses when file transfers occurred.<br>• Associated timestamps<br>• Web documents and image files from the HTTP sites. |

Taken together, it can be seen that the findings echo the trend of users storing their data in the cloud. Unsurprisingly, AIM 7 uses encrypted channels to allow users to communicate securely and privately. However, this complicates forensic investigations. On 16 October 2014, for example, the Federal Bureau of Investigation's director remarked that 'going dark' (i.e., law enforcement not being able to access evidential data due to technological measures in place to protect the security and privacy of technology users) *"will have very serious consequences for law enforcement and national security agencies at all levels [as] Sophisticated criminals will come to count on these means of evading detection"* (FBI, 2014).

To keep pace with technological advances, future work would include investigating the newer versions and AIM for other platforms to have an up-to-date forensic understanding of the artefacts that can be used to inform investigations.

## Acknowledgements

The authors would like to thank the anonymous reviewers for providing constructive and generous feedback. Despite their invaluable assistance, any errors remaining in this paper are solely attributed to the author.